# Dust-Scattered Ultraviolet Halos around Bright Stars

**Jayant Murthy[1*] & Richard Conn Henry[2]**

[1]Indian Institute of Astrophysics, Bengalooru 560 034, India, [2]Dept. of Physics and Astronomy, The Johns Hopkins University, Baltimore, MD. 21218, USA

*Correspondence address: jmurthy@yahoo.com

## *Abstract*

We have discovered ultraviolet halos extending as far as 5° around four (of six) bright UV stars using data from the *GALEX* satellite. These halos are due to the scattering of the starlight from nearby thin, foreground dust clouds. We have placed limits of 0.58 ± 0.12 and 0.72 ± 0.06 on the phase function asymmetry factor (*g*) in the FUV (1521 Å) and NUV (2320 Å) bands, respectively. We suggest that these halos are a common feature around bright stars and may be used to explore the scattering function of interstellar grains at small angles.

## *Introduction*

A single bright star may illuminate the interstellar dust to a considerable distance and, in point of fact, halos around two bright stars (Spica and Achernar) are clearly visible in the all-sky ultraviolet (UV) maps of Murthy, Henry, & Sujatha (2010). These halos represent the scattered light from a small amount of interstellar dust located between us and the star, but not physically associated with the star, and their analysis has the potential to unambiguously determine the scattering properties of the grains. Until now, most determinations of the optical constants of the grains have been through observations of either reflection nebulae or the diffuse UV background but the interpretation of both has been fraught with difficulty and often not yielded unique results (Mathis, Whitney, & Wood 2002). Nevertheless, there is a consensus that the grains are moderately reflective and strongly forward scattering in the UV with an albedo (*a*) of 0.4 - 0.6 and a phase function asymmetry factor (*g*) of about 0.6 (see review by Draine 2003).

We have used observations from the *Galaxy Evolution Explorer* (*GALEX*) to search for dust-scattered halos near UV bright stars. *GALEX (*described by Martin et al. 2005) has observed close to 80% of the sky in two wavelength bands (FUV: 1350 - 1700 Å and NUV: 1700 - 3200 Å) with observations of diffuse radiation at all latitudes outside of the Galactic plane and away from bright stars, for safety reasons. Thus, of the 50 brightest stars in the UV sky, *GALEX* only observed near (within 5°) 6, including Spica and Achernar, the third and fifth brightest stars in the sky, respectively. Of these 6 stars, we have found stellar halos around 4 which we identify as scattering from dust relatively near to and in front of the star. Unlike classical reflection nebulae where the star is embedded in an optically thick cloud, these halos are due to the single scattering of the starlight from a thin layer of dust conveniently between us and the star. The interpretation of the results is therefore much easier and provides, in principle, an unambiguous measurement of the optical constants of the interstellar dust grains, particularly the phase function asymmetry factor (*g*).

## *Observations and Data Analysis*

As mentioned above, *GALEX* observed near only 6 of the 50 brightest UV stars (Table 1). Most of these observations were part of the All-Sky Imaging Survey (AIS) with typical exposure times of about 100 seconds. Although there were a few longer duration observations, these comprised several visits over widely separated dates and, for the sake of consistency, we left them out of this sample. Three of the six stars (Spica, Achernar, and Algenib) were well covered with many nearby observations while the remaining three had only a handful of nearby observations with a concomitant loss in the signal-to-noise ratio.

Our data analysis procedure is a modified version of that used by Sujatha et al. (2010). We first removed all the point sources from the *GALEX* FUV and NUV images using the merged source catalog provided with each observation, leaving only the diffuse emission. A significant fraction of this emission is due to airglow, primarily resonantly scattered solar photons from O I in the Earth's atmosphere, and zodiacal light from sunlight scattered by interplanetary dust grains. We estimated the airglow using empirical relations between the strength of the airglow in each band and the level of the solar radiation found by Sujatha et al. (2010). The zodiacal light contributes only to the NUV band and was estimated by scaling from optical observations tabulated by Leinert et al. (1998), assuming no color correction relative to the solar spectrum. The foreground emission amounted to about 200 - 400 photons cm$^{-2}$ s$^{-1}$ sr$^{-1}$ Å$^{-1}$ in the FUV channel and 400 - 600 photons cm$^{-2}$ s$^{-1}$ sr$^{-1}$ Å$^{-1}$ in the NUV which we subtracted from each observation, with an uncertainty of about 50 photons cm$^{-2}$ s$^{-1}$ sr$^{-1}$ Å$^{-1}$ in the level of the subtracted light.

We averaged the remaining emission into annular rings of width 0.05° around the star along with the average flux in the *Infrared Astronomy Satellite (IRAS)* 60 and 100 micron bands (Fig. 1). UV enhancements are seen around 4 of the 6 stars observed with most also showing an enhancement in the 60/100 μm ratio. The emission far from the star, in both the UV and the IR, is due to the general diffuse radiation field (Murthy et al. 2010) but, as these are bright early-type stars, the stellar sphere of influence extends far into the surrounding interstellar medium. The ratio between the diffuse UV flux and that of the star ranges from about 6 x 10$^{-8}$ in the case of Achernar to 6 x 10$^{-7}$ for Mirzam (Fig. 2). We did not observe an enhancement near Adhara, perhaps because there were no observations nearer than 3° from the star, nor was there an enhancement around Sirius, despite having data as close as 2° to the star. As a consequence, we have proceeded with the assumption that instrumental scattering is unimportant at these angular distances even from bright sources.

## *Modeling*

The geometry in our observed fields is particularly simple. A bright star dominates the nearby radiation field and its light is forward scattered by a thin layer of dust between us and the star. We have used Kurucz (Kurucz 1979) model spectra convolved with the instrumental calibration to calculate the count rate in the *GALEX* FUV and NUV bands (Table 1) for each star. This light was scattered from a plane-parallel sheet of dust in front of the star where the amount of dust in each pixel was given by the extinction maps of Schlegel et al. (1998). The scattering was assumed to be single scattering, as the dust is optically thin, and was governed by the Henyey-Greenstein scattering function (Henyey & Greenstein 1941), an analytical approximation to Mie scattering from spherical dust grains with two parameters: the albedo (*a*) and the phase function asymmetry factor ($g = <\cos\theta>$).

The level of the scattered radiation is directly proportional to *a* and to the optical depth in the optically thin limit but, as noted by Lee et al. (2008), there is a more complex trade-off between the distance of the cloud from the star and *g*. This is illustrated in Fig. 3 where we have calculated the expected emission for a plane parallel dust sheet in front of Spica, assuming an albedo of 0.1. The closer the dust is to the star, the stronger the halo will be, even with isotropic scattering ($g = 0$). On the other hand, if the dust is far from the star, halos will only be seen if the grains are strongly forward scattering. The slope flattens out at greater angular separations from the star in all cases and, as a result, there is some ambiguity in modeling the data at the angles where we have observations (> 2°). Observations closer to the star would help to remove this degeneracy but are difficult because of the increasing brightness of the field close to the star.

Fortunately, we have an additional constraint on the location of the dust responsible for most of the observed UV radiation through the observed *IRAS* 60/100 μm ratio. Dust near the star will be heated by the stellar radiation field and the ratio will increase if the cloud is closer to the star. Draine & Li (2007) have tabulated the expected emission in the two *IRAS* bands as a function of the radiation field for a mixture of silicate and carbonaceous grains and we have used this to estimate the distance between the star and the dust (Table 2). We finally generated model fluxes for each pixel for a range of parameter values and compared with the observational data using standard $\chi^2$ techniques (Lampton, Margon, & Bowyer 1976). Rather than comparing the model values with the annular averages of Fig. 1, we averaged the *GALEX* data into 0.25 square degree bins, which gave us a finer granulation in the sky. These data (Table 3) are plotted against representative model values in Fig. 4. We have found an empirical scatter in our data of about 300 photons $cm^{-2}\ s^{-1}\ sr^{-1}\ Å^{-1}$, much larger than the intrinsic photon noise of a few photons $cm^{-2}\ s^{-1}\ sr^{-1}\ Å^{-1}$. About 50 photons $cm^{-2}\ s^{-1}\ sr^{-1}\ Å^{-1}$ of this is due to the subtraction of the foreground airglow and zodiacal light (Sujatha et al. 2010) but more important is the modeling uncertainty due to the difficulty in estimating the amount of dust in the scattering layer, particularly at the low column densities applicable in this work. In addition, the local interstellar medium is complex (Redfield & Linsky 2008) and it may well be that the dust is split between two or more clouds with the bulk of the UV and IR emission coming from the cloud nearest the star. We will discuss the effects of this assumption further in the next section.

## *Results*

Interestingly, dust-scattered halos were seen around 4 of the 6 stars studied here. Of the other 2 stars, Sirius is too close to the Sun to have any foreground dust while our models indicate that any enhancement around Adhara would only extend out about 3° from the star, where there were no observations. Similar halos were observed by Lee et al. (2008) around stars in front of the Ophiuchus molecular cloud using the *SPEAR* UV spectrograph and it is quite likely that such enhancements exist around the majority of bright stars, although they may not always be observable. The brightness of the halo will be governed by a number of parameters including the optical constants of the grains, the distance between the star and the dust, the intrinsic luminosity of the star and the total amount of dust in the scattering layer while the angular dependence is determined by the distance of the dust to the star and the scattering function. As discussed above, we have used the 60/100 μm ratio to infer a distance of the dust from each star, which we find to range from 3 to 10 pc resulting in scattering angles (between the starlight and the scattered radiation) of between 20° and 50°.

We have plotted the allowed ranges for the optical constants for each of the 4 stars with halos in Fig. 5 with representative model fits in Fig. 4. The brightest of our halos is around Spica from a dust cloud noted by Zagury et al. (1998) which we place at a distance of 3 pc from the star using the 60/100 μm ratio (Table 2). We find formal 1σ bounds of $0.58 \pm 0.12$ on the phase function asymmetry factor (*g*) in the FUV (Fig. 5a) and and $0.72 \pm 0.06$ in the NUV (Fig 5b), in agreement with most determinations in the literature (Table 4). The corresponding limits on the albedo are $0.10 \pm 0.05$ in the FUV and $0.26 \pm 0.10$ in the NUV.

These low albedos are a direct consequence of our assumption that the dust is concentrated in a single plane-parallel sheet. Rather, it is likely that the dust is distributed along the line of sight in multiple components. It may not be possible to find a unique solution with the available information but we have explored one possibility: that of two plane-parallel clouds at different distances from the star. One such example is shown in Fig. 6 where we have divided the dust between two clouds such that 20% of the total matter is in a cloud at 2 pc from Spica with the remainder in a cloud at a distance of 30 pc from the star. The albedo of the dust in both clouds is fixed at 0.4 with *g* at 0.6. Most of the angular dependence for moderate values of *g* is contained in the cloud nearer to the star with the more distant cloud providing a flat offset. We have found that the shape of the curve is difficult to fit with *g* values differing by much from 0.6 but are hampered by not having observations nearer to the star.

The diffuse emission is too intense to observe much closer to any of these stars with GALEX; however, the emission around Spica is also visible as a bright spot in the SPEAR/FIMS maps (Edelstein et al. 2006). Although Park et al. (2010) focused on the Hα emission around Spica, a strong continuum emission is apparent in their spectrum with levels of 31,000 photons cm$^{-2}$ s$^{-1}$ sr$^{-1}$ Å$^{-1}$ and 13,000 photons cm$^{-2}$ s$^{-1}$ sr$^{-1}$ Å$^{-1}$ at angular distances of 0° - 1° and 1° - 2°, respectively (Park 2010). These are plotted in Fig. 7 along with our GALEX data and three models for different values of *g*. It is difficult to fit the SPEAR data in conjunction with the GALEX data, perhaps because the SPEAR data are averaged over large angular ranges with a wide variation in flux but, certainly, very large values of *g* (*g* > 0.7) are ruled out.

## *Conclusions*

We have observed dust halos around 4 out of 6 stars, extending as far as 5º from the star. Of the remaining two stars, Sirius is close enough to the Sun (8 pc) that there is not enough foreground dust to scatter the light while we believe that observations nearer to Adhara would have detected a halo. Halos were also observed around many stars in Ophiuchus by Lee et al. (2008) and we suggest that such halos, where starlight from bright stars is scattered by foreground dust, are a common occurrence, although their detection may prove to be challenging closer to the star, because of instrumental scattering.

We have placed 1σ limits on the phase function asymmetry factor (*g*) of 0.58 ± 0.12 in the FUV and 0.72 ± 0.06 in the NUV but were unable to constrain the albedo because of an uncertain dust distribution. These observations have shown their potential in understanding the scattering function of the dust grains but are frustratingly limited by not observing close enough to the stars. There is a limited window of opportunity to obtain new *GALEX* observations but we will also explore other opportunities with other instruments. We are also attempting better models combining infrared data available from *IRAS* and other instruments to further constrain the dust properties.

Acknowledgments: We thank an anonymous referee for inducing us to clarify the geometrical uncertainty in the model. Greg Schwarz helped us greatly in publishing our data. The research leading to these results received funding from the Indian Space Research Organization through the Space Science Office and the National Aeronautics and Space Administration through the Maryland Space Grant program. This research is based on observations made with the NASA's GALEX program, obtained from the data archive at the Space Telescope Science Institute. STScI is operated by the Association of Universities for Research in Astronomy, Inc. under NASA contract NAS 5-26555.

## Figures

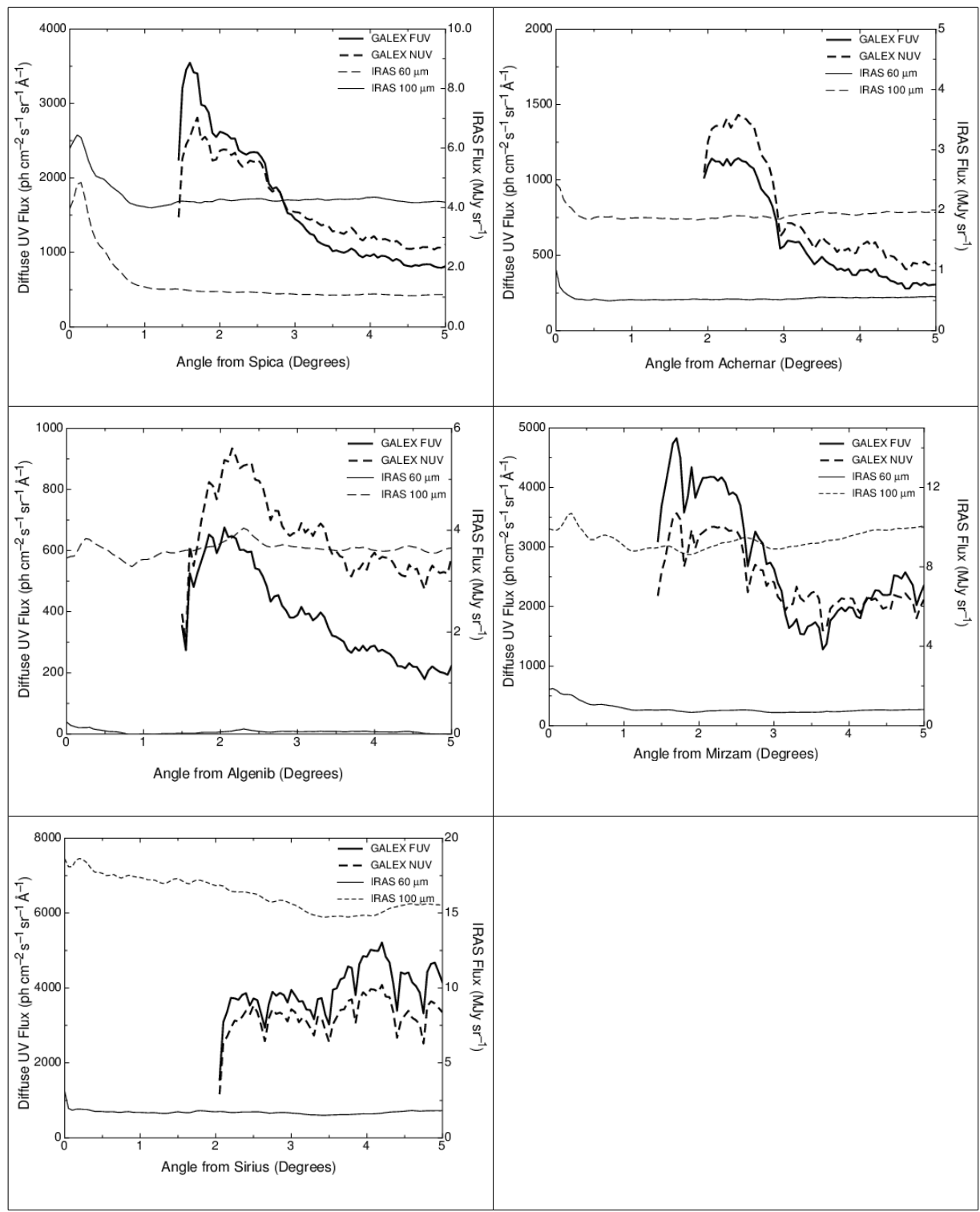

Fig. 1 - Angular dependence of UV emission for Spica (Fig. 1a), Achernar (Fig. 1b), Algenib (Fig. 1c), Mirzam (Fig. 1d), Adhara (Fig. 1e), and Sirius (Fig. 1f). The downturn in the profiles at small angles is an artifact of the averaging process. The IR emission is superimposed on each figure.

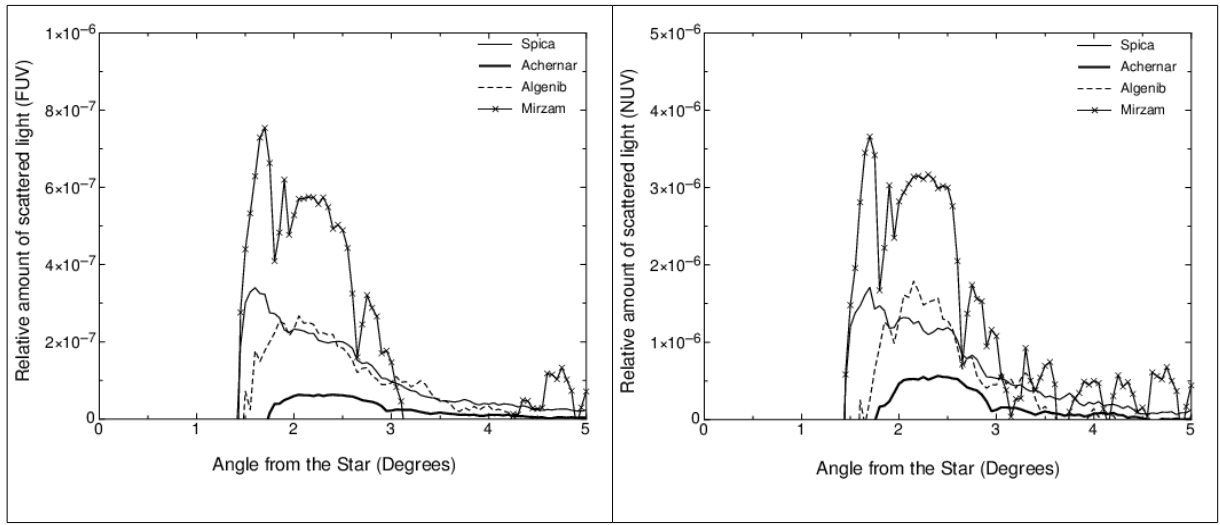

Fig. 2 - Ratio between diffuse flux and stellar flux for the four stars with halos for FUV (Fig. 2a) and NUV (Fig. 2b). A background has been subtracted from the diffuse flux such that the level is zero at large angles. The drop in the profiles near the star is an artifact of the averaging process.

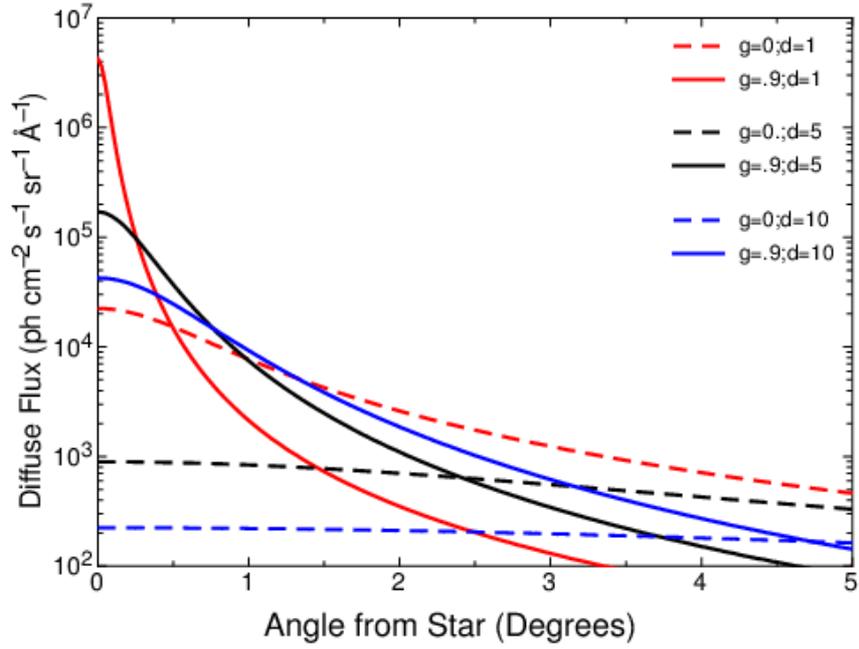

Fig. 3 – Dependence of scattering on *g* and distance from the star (Spica in this example). Halos are more prominent when the scattering layer is near the star and the grains are forward scattering. Observations closer to the star provide a better discriminator of the optical constants.

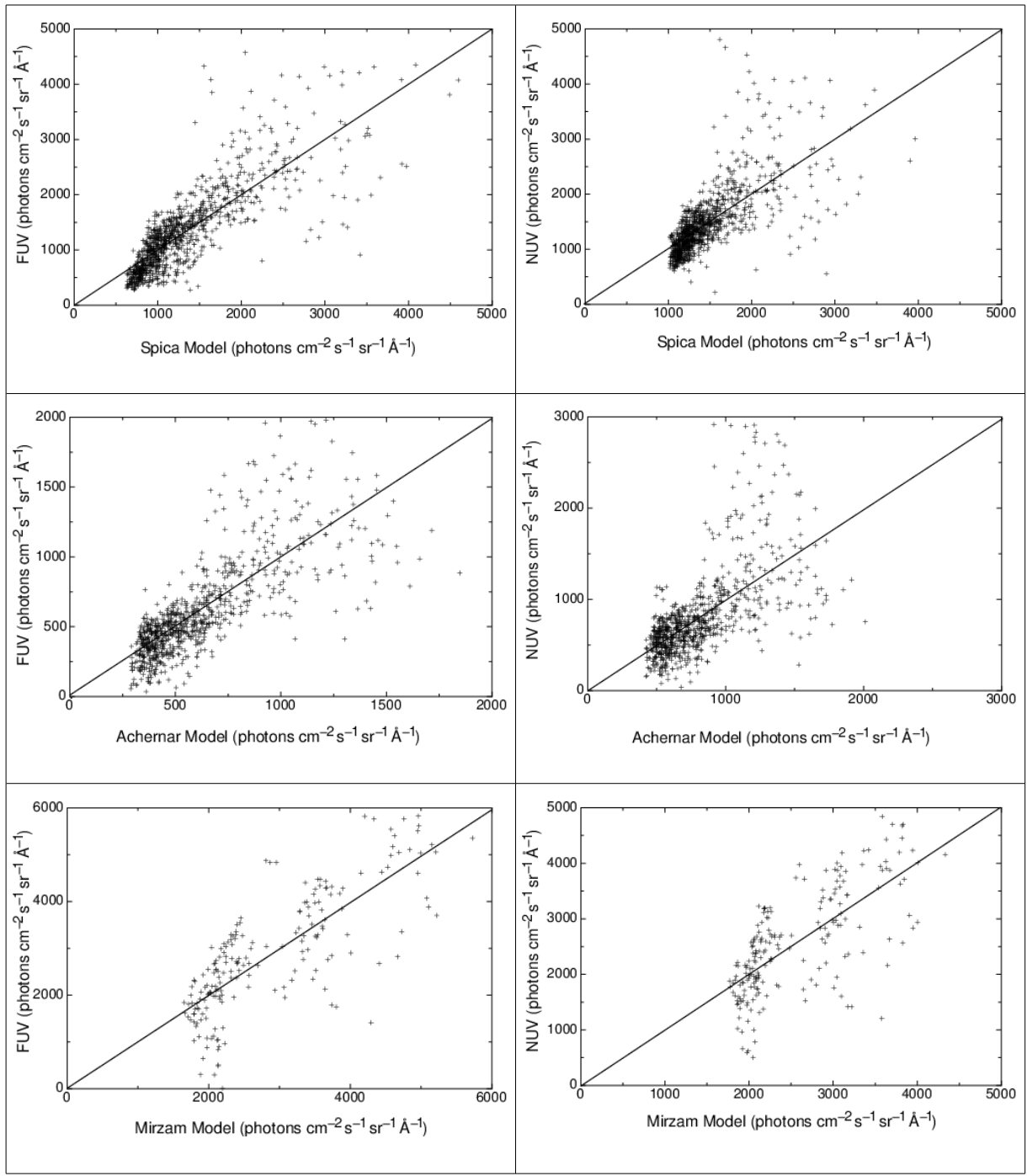
13

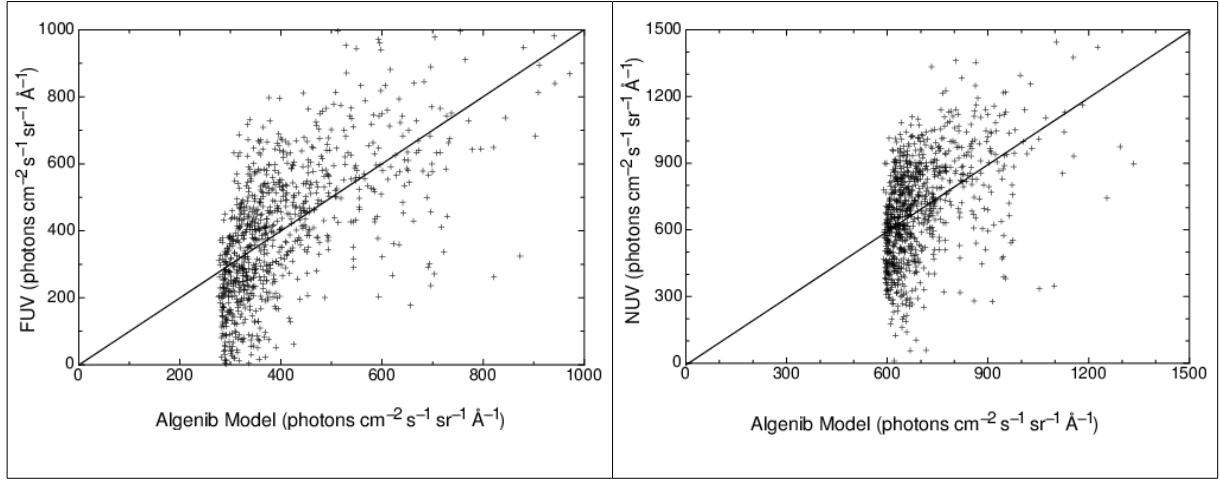

Fig. 4 – Model fits for Spica FUV (Fig. 4a: minimum $\chi^2$ = 1.34) and NUV (Fig. 4b: minimum $\chi^2$ = 1.16); Achernar FUV (Fig. 4c: minimum $\chi^2$ = 0.915) and NUV (Fig. 4d: minimum $\chi^2$ = 1.77); Mirzam FUV (Fig. 4e: minimum $\chi^2$ = 11.79) and NUV (Fig. 4f): minimum $\chi^2$ = 8.43; and Algenib FUV (Fig. 4g: minimum $\chi^2$ = 0.348) and NUV (Fig. 4h: minimum $\chi^2$ = 0.656). Airglow and zodiacal light have been subtracted from the data values and we have assumed a scatter of 300 photons cm$^{-2}$ s$^{-1}$ sr$^{-1}$ Å$^{-1}$ to account for both experimental and modeling uncertainties. An x=y line has been added to each plot for reference.

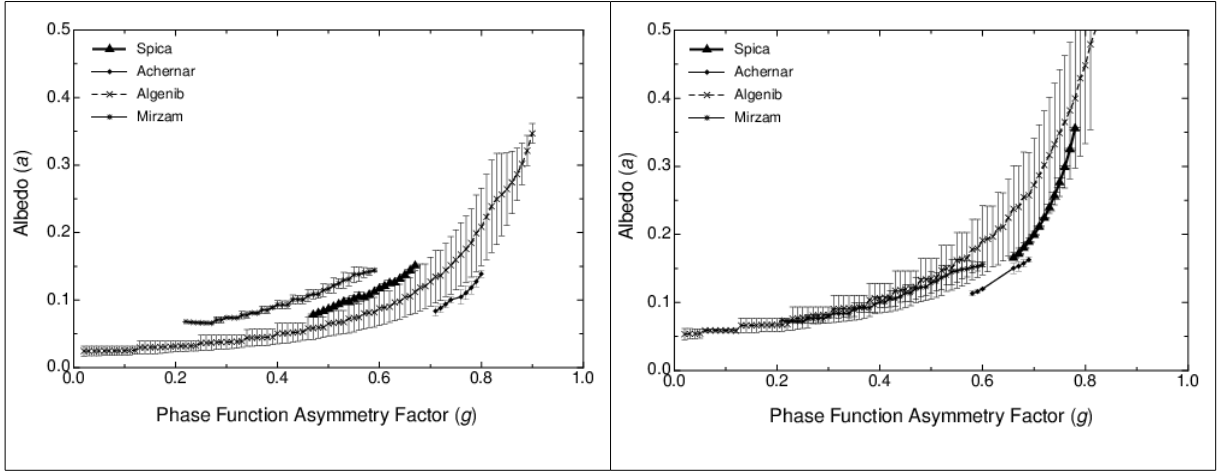

Fig. 5 – Allowed phase space for *a* and *g* for the FUV (Fig. 5a) and NUV (Fig. 5b).

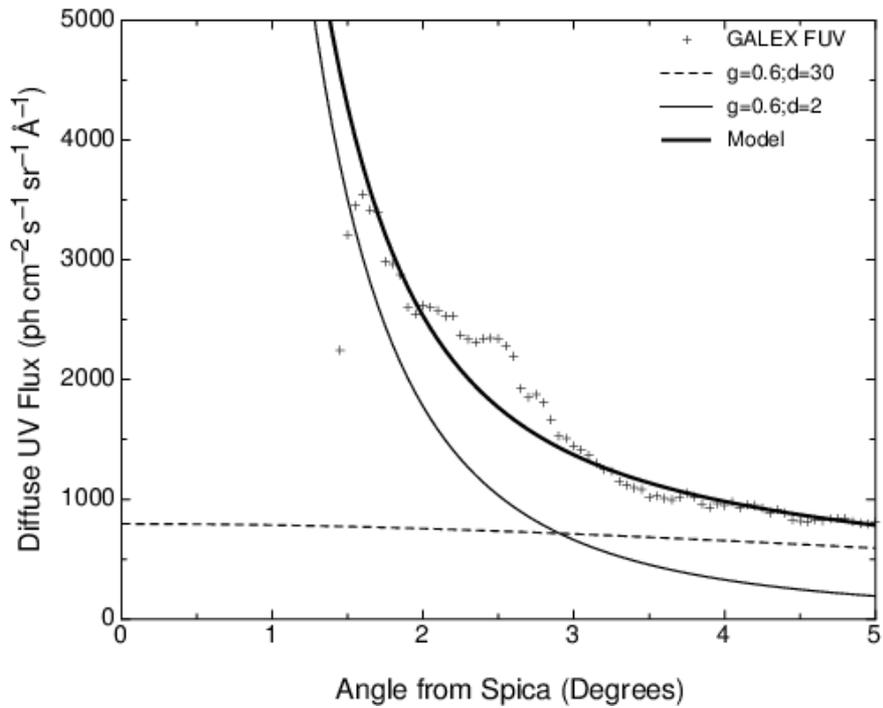

Fig. 6 – Model fit with two plane-parallel clouds at distances of 2 and 30 pc, respectively, from Spica with 20% of the total column density in the first cloud and 80% in the second. The optical constants are fixed with *a* = 0.5 and *g* = 0.6. The *SPEAR* data points represent integrations from 0° - 1° and 1° - 2° where the flux changes rapidly.

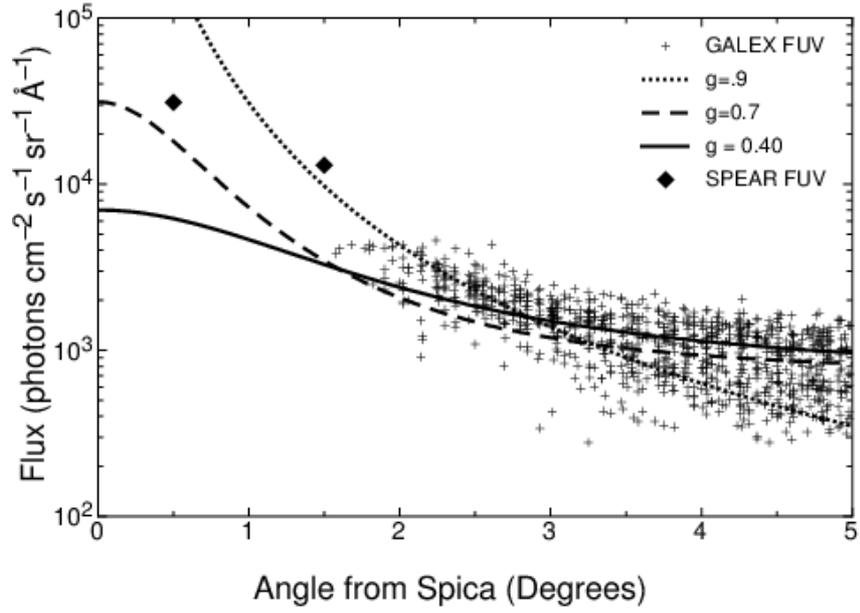

Fig. 7 – FUV flux with angle (+) for Spica. Three different profiles are plotted for different values of *g* for a dust cloud 3 pc from the star. The two SPEAR points (Park 2010) are plotted as diamonds.

**Table 1**

Observation Log

| Star | Sp. Type | l | b | d[a] | N(HI)[b] | FUV[c] | NUV[c] | No.[d] |
|---|---|---|---|---|---|---|---|---|
| Spica (α Vir) | B1 III | 316.11 | 50.85 | 80 | 19.0 | 4204 | 2367 | 90 |
| Achernar (α Eri) | B3 Vpe | 290.84 | -58.79 | 44 | 18.8 | 6912 | 3892 | 56 |
| Algenib (γ Peg) | B2IV | 109.43 | -46.68 | 103 | 20.0 | 808 | 455 | 86 |
| Sirius (α CMa) | A1V | 227.23 | -8.89 | 2.6 | 18.0 | 1549 | 1630 | 4 |
| Mirzam (β CMa) | B1II-III | 226.06 | -14.27 | 150 | 18.4 | 1768 | 995 | 4 |
| Adhara (ε CMa) | B2 Iab | 239.83 | -11.33 | 132 | 18.0 | 2751 | 1549 | 7 |

[a]Distance in pc to star.
[b]Column density to star (Shull & van Steenberg 1985)
[c]Predicted count rate in *GALEX* bands in counts s$^{-1}$.
[d]Number of observations within 5° of the star.

**Table 2**

Distance of Scattering Layer from Star

| Star | 60/100 μm Ratio | Distance of dust (pc) |
|---|---|---|
| Spica | 0.75 | 3 |
| Achernar | 0.4 | 3 |
| Mirzam | 0.15 | >10 |
| Adhara | 0.3 | 8 |
| Algenib | - | - |

**Table 3**

GALEX FUV and NUV around Stars

| Name[a] | GLON[b] | GLAT[c] | Angle[d] | FUV[e] | NUV[f] | E(B-V)[g] |
|---|---|---|---|---|---|---|
| Spica | 313.71 | 51.32 | 1.58 | 3809.73 | 2607.71 | 0.042 |
| Spica | 317.74 | 52.08 | 1.60 | 4073.17 | 3005.78 | 0.044 |
| Spica | 313.70 | 51.57 | 1.68 | 4311.21 | 3185.18 | 0.037 |
| Spica | 318.15 | 52.08 | 1.77 | 4079.03 | 3618.86 | 0.046 |
| Spica | 317.75 | 52.33 | 1.80 | 4349.50 | 3893.84 | 0.050 |
| Spica | 313.69 | 51.82 | 1.80 | 3983.68 | 3569.00 | 0.038 |
| Spica | 315.29 | 52.59 | 1.82 | 2495.77 | 2348.72 | 0.029 |
| Spica | 313.31 | 51.31 | 1.82 | 4226.48 | 3409.33 | 0.039 |
| Spica | 313.30 | 51.56 | 1.90 | 4140.50 | 3589.98 | 0.035 |
| Spica | 313.39 | 50.06 | 1.90 | 3074.67 | 2271.59 | 0.048 |

Notes

[a] The name of the illuminating star.
[b] Galactic longitude of the pixel.
[c] Galactic latitude of the pixel.
[d] Angle in degrees from the illuminating star.
[e] Flux in the *GALEX* FUV band in units of photons cm$^{-2}$ s$^{-1}$ sr$^{-1}$ Å$^{-1}$.
[f] Flux in the *GALEX* NUV band in units of photons cm$^{-2}$ s$^{-1}$ sr$^{-1}$ Å$^{-1}$.
[g] E(B-V) from Schlegel et al. (1998).
Only a portion of this table is shown here to demonstrate its form and content. A machine-readable version of the full table is available at ApJ.

**Table 4**

Optical Constants

| λ (Å) | a | Δa | g | Δg | Region | Reference |
|---|---|---|---|---|---|---|
| FUV | | | | | | |
| 1500 | 0.42 | - | 0.44 | - | DGL | Wright (1992) |
| 1500 | 0.5 | - | 0.9 | - | DGL | Witt & Petersohn (1994) |
| 1500 | 0.44 | 0.22 | - | - | DGL | Murthy & Henry (1995) |
| 1520 | 0.36 | 0.2 | 0.52 | 0.22 | DGL | Lee et al. (2008) |
| 1521 | 0.4 | 0.01 | 0.7 | 0.01 | DGL | Sujatha et al. (2009) |
| 1521 | 0.32 | 0.09 | 0.51 | 0.19 | DGL | Sujatha et al. (2010) |
| 1521 | 0.12 | 0.04 | 0.57 | 0.1 | Spica | This work |
| 1521 | 0.11 | 0.03 | 0.75 | 0.05 | Achernar | This work |
| 1521 | 0.1 | 0.04 | 0.4 | 0.18 | Mirzam | This work |
| 1521 | 0.18 | 0.16 | 0.45 | 0.4 | Algenib | This work |
| 1525 | 0.77 | 0.05 | 0.66 | 0.05 | IC 435 | Calzetti et al. (1995) |
| 1550 | 0.6 | 0.05 | 0.7 | 0.1 | DGL | Lillie & Witt (1976) |
| 1550 | 0.5 | - | 0.5 | - | DGL | Morgan et al. (1976) |
| NUV | | | | | | |
| 2275 | 0.39 | 0.05 | 0.7 | 0.05 | IC 435 | Calzetti et al. (1995) |
| 2320 | 0.4 | - | 0.7 | - | DGL | Sujatha et al. (2009) |
| 2320 | 0.45 | 0.08 | 0.56 | 0.1 | DGL | Sujatha et al. (2010) |
| 2320 | 0.26 | 0.09 | 0.73 | 0.04 | Spica | This work |
| 2320 | 0.14 | 0.03 | 0.63 | 0.06 | Achernar | This work |
| 2320 | 0.11 | 0.04 | 0.4 | 0.2 | Mirzam | This work |
| 2320 | 0.32 | 0.3 | 0.45 | 0.4 | Algenib | This work |
| 2325 | 0.38 | 0.05 | 0.78 | 0.05 | IC 435 | Calzetti et al. (1995) |
| 2350 | 0.58 | 0.15 | -1 | 0 | DGL | Morgan et al. (1976) |